\documentstyle[12pt]{article}
\newcommand{\be}{\begin{equation}}
\newcommand{\ee}{\end{equation}}
\newcommand{\bea}{\begin{eqnarray}}
\newcommand{\eea}{\end{eqnarray}}

\newcommand{\bb}{\bibitem}
\newcommand{\pa}{\partial}
\newcommand{\beq}{\begin{equation}}
\newcommand{\eeq}{\end{equation}}
\newcommand{\ba}{\begin{eqnarray}}
\newcommand{\ea}{\end{eqnarray}}

\begin{document}
\begin{titlepage}
\begin{center}
\LARGE

{Holographic Principle and  AdS/CFT Correspondence\footnote{Talk
     presented at the International Workshop ``Supersymmetries and
     Quantum Symmetries'', held at 
     JINR, Dubna,  July 26-31, 1999}
     }
\vspace{.5cm}
\large

Victor O. Rivelles 

\vspace{.5cm}{\it Instituto de F\'\i sica, Universidade de S\~ao 
Paulo} \\ {\it C.Postal 66318, 05315-970, S.Paulo, SP, Brazil} \\ {
rivelles@fma.if.usp.br}

\vspace{.5cm}

\end{center}
\vspace{1cm}

\begin{abstract}
According to the holographic principle all information in the bulk of
a space is coded at its border. We will check this
statement in three situations involving the AdS/CFT correspondence. 
There is a well known equivalence between the 
Maxwell-Chern-Simons theory and the self-dual model in 3
dimensions when the parameters of both theories are related in a given
way. We will show that when this relation holds the corresponding
CFT's at the border are the same. Then we will study scalar fields.
There are two quantum theories for the scalar field in AdS space. The
usual prescription of the AdS/CFT correspondence which takes Dirichlet
boundary conditions at the border corresponds to one of the quantum
theories. We will show that changing boundary conditions will allow us
to get the other quantum theory. Finally we consider an Abelian gauge
theory in AdS. We will show that the corresponding CFT is independent
of the gauge choice and that the gauge dependence stays only in the
contact terms at the border. 
\end{abstract}
\end{titlepage}

\section{INTRODUCTION}

The holographic principle states that a quantum theory with gravity
must be describable by a boundary theory \cite{holographic}. Of course
this raises the question on how the detailed information in the bulk
can be completely stored at the border and this surely deserves a deep
investigation. A possible way to investigate this connection is
through the AdS/CFT 
correspondence \cite{maldacena}. It says that the large $N$ limit of a
certain conformal field theory (CFT) in a $d$-dimensional Minkowski
space can be 
described by string/M-theory on $AdS_{d+1} \times K$ where $K$ is a
suitable compact space. The precise relation between both theories is
given by \cite{ads/cft}
\be
\label{101}
Z_{AdS}[\phi_0]\,=\,\int_{\phi_0}\,{\cal D}\phi\,\exp(-
I[\phi])\,\equiv\,Z_{CFT}[\phi_0]\,=\, \langle \exp \left(\int_{\pa \Omega}\,
d^d x\,{\cal O}\phi_0\right)\rangle\,\,\,,
\ee
where $\phi_0$ is the value taken by $\phi$ at the border. On the
right handed side $\phi_0$ is the external current coupling to the
operator ${\cal O}$ in the boundary CFT. Hence the partition
function in $AdS_{d+1}$ allow us to obtain the correlation functions
of the boundary CFT. In this sense the AdS/CFT correspondence is a
realization of the holographic principle. 

The AdS/CFT correspondence in the form Eq.(\ref{101}) has been studied
in several situations \cite{review}. We will analyze it the three
different situations. In the first case we will concentrate in
$d=3$. As it is well known there is a equivalence \cite{equiv} between the
Maxwell-Chern-Simons theory \cite{mcs} and the self-dual model
\cite{sd} in Minkowski space. Both theories are equivalent when the
parameters have a precise relation. We have been able to show that this
equivalence also holds in AdS space \cite{cs}. We can then use the
AdS/CFT correspondence to compute the two point functions at the
border. When the parameters are chosen so that the equivalence holds
we can show that the corresponding CFT's at the border are the same
\cite{cs}. This shows that a relationship between two quantum field
theories in AdS is directly reflected in the corresponding CFT's at
the border. This is discussed in Section \ref{s2}. 

The second case we study regards a scalar field. There are two
distinct quantum field theories for the scalar field in AdS. They
depend on which energy-momentum tensor is chosen \cite{scalar}. Using
the AdS/CFT correspondence in the form Eq.(\ref{101}), which uses
Dirichlet boundary conditions, one of the two quantum theories for the
scalar field is reproduced. The other one can be obtained by
considering a different boundary condition
\cite{revisited}. Therefore, as expected, 
both quantum theories can be obtained using the AdS/CFT
correspondence.  This will
be presented in Section \ref{s3}. 

Finally we will consider the gauge dependence in the AdS/CFT
context. We take an Abelian gauge theory in arbitrary gauge in an 
AdS background. We will show that the conformal correlators do not
depend on the gauge parameter. However the holographic principle
asserts that the information about the gauge degrees of freedom must
survive at the border. We indeed find a gauge dependence only in the
contact terms which usually are thrown away \cite{gauge}. This will be
the subject of Section \ref{s4}.

\section{CHERN-SIMONS THEORIES}
\label{s2}

Since we are going to consider the Euclidean version of $AdS _{3}$ we
start with the Euclidean signature action for the Proca-Chern-Simons
theory which is given by 
\beq
I_{PCS} =  \int d^3x \, \sqrt{g} \left( \frac{1}{8}F
  _{\mu\nu}F^{\mu\nu} + \frac{1}{4}m^{2}A _{\mu}A^{\mu} + 
\frac{1}{\sqrt{g}}\frac{i\mu}{8}\epsilon^{\mu\nu\alpha}F
_{\mu\nu}A_{\alpha}\;+\;c.c. \right),
\label{1}
\eeq 
where $F _{\mu\nu} = \partial _{\mu}A _{\nu}-\partial _{\nu}A
_{\mu}$ and $\epsilon^{\mu\nu\alpha}$ is the Levi--Civita tensor
density with $\epsilon^{012}=1$. The field equations which follow from
Eq.(\ref{1}) are 
\beq
\nabla_{\mu} F^{\mu\nu} - m^2 A^{\nu} - i \mu \frac{1}{\sqrt{g}} 
\epsilon^{\nu\alpha\beta}\partial_{\alpha}A_{\beta} = 0,
\label{11}
\eeq
and they can be manipulated to give 
\beq
\left( \nabla^2 - m_+^2 - \frac{R}{3} \right) \left( \nabla^2 - m_-^2
  - \frac{R}{3} \right) A^\mu = 0,
\label{14}
\eeq
where 
\beq
m^2_\pm(m,\mu) = \left[ \left( m^2 + \frac{\mu^2}{4} \right)^\frac{1}{2}
  \pm \frac{\mu}{2} \right]^2,
\label{masses}
\eeq
and $R=-6$ is the AdS radius. We notice that the
solutions of Eq.(\ref{14}) must satisfy 
\beq
\left( \nabla^2 - m_+^2 - \frac{R}{3} \right) A^\mu = 0, 
\label{15}
\eeq
or
\beq
\left( \nabla^2 - m_-^2 - \frac{R}{3} \right) A^\mu = 0.
\label{16}
\eeq
Therefore the general solution of Eq.(\ref{14}) is a superposition of
solutions of the Proca theory with masses $m_+$ and $m_-$. In the flat
space limit we recover the fact that the Proca-Chern-Simons theory
describes two massive excitations. The Proca
theory in AdS space has been analyzed in detail in \cite{Viswa}. 

We can now calculate the bulk to bulk propagator. The details of the
calculation can be found in \cite{cs}. Then we can evaluate the
classical action near the boundary surface using the
action Eq.(\ref{1}). After an 
integration by parts and using the equations of motion we find that
there is only a contribution from the boundary 
\beq
I_{PCS} =  \frac{1}{4} \int d^3x \; \partial_\mu ( \sqrt{g}
F^{\mu\nu} A_\nu )\;+\;c.c., 
\label{boundary}
\eeq
which evaluated on the near boundary surface gives
\beq
I _{PCS} = -\frac{1}{4}\int d^2x \; \epsilon^{-2}{\tilde A
_{\epsilon,i}}\left(
-{\tilde A _{\epsilon,i}} + \epsilon{\tilde F
_{\epsilon,0i}}\right)\;+\;c.c..
\label{115}
\eeq 
Here $\tilde A _{\epsilon,i}$ is the value of the field at the near
boundary surface. When inserting the propagator to evaluate $I_{PCS}$
we find that the result is divergent in the limit $\epsilon
\rightarrow 0$ so that a regularization has to be introduced. In order
to have a finite action we take the limit 
\beq
\lim _{\epsilon\rightarrow 0}\epsilon^{m_{-}(m,\;\mid\mu\mid)
- 1}{\tilde
A_{\epsilon,i}}(\vec{x}) = {\tilde A_{0,i}}(\vec{x}).
\label{133}
\eeq

Then we use the AdS/CFT correspondence in the form
\beq
\exp\left( -I _{AdS}\right) \equiv \left<\exp\left(\int d^2x \;  
J_{i}(\vec{x}) \; A _{0,i}(\vec{x})\right)\right>,
\label{134}
\eeq
and we find the two point function 
\beq
\left <J _{i}^{PCS}(\vec{x}) \; J_{j}^{PCS}(\vec{y})\right> =
{\tilde c _{PCS}}{\tilde \Delta
_{PCS}}\left(
\delta_{ij}-2\frac{(x-y)_{i}(x-y)_{j}}{\mid\vec{x}-\vec{y}\mid^{2}}\right)
\mid\vec{x} - \vec{y}\mid^{-2{\tilde 
\Delta_{PCS}}},
\label{139}
\eeq
where $
{\tilde \Delta_{PCS}} = {\tilde \Delta}_{-}(m,\;\mid\mu\mid)$, 
and $ {\tilde c _{PCS}} = {\tilde c _{-}}(m,\;\mid\mu\mid)$, 
so that $J_{i}^{PCS}$ has conformal dimension ${\tilde \Delta_{PCS}}$. 
It is important to note that the identification Eq.(\ref{133}) agrees 
with the requirement 
that the isometries of $AdS_{3}$ correspond to the conformal isometries in
$CFT_{2}$.

In order to get the boundary CFT associated to the Maxwell-Chern-Simons
theory we take $m = 0$ in Eq.(\ref{139}), which gives
\beq
\left <J _{i}^{MCS}(\vec{x}) \; J _{j}^{MCS}(\vec{y})\right> =
{\tilde c _{MCS}}{\tilde \Delta
_{MCS}}\left( \delta
_{ij} - 2\frac{(x-y) _{i}(x-y) _{j}}{\mid\vec{x} -
\vec{y}\mid^{2}}\right)\mid\vec{x} - \vec{y}\mid^{-2{\tilde \Delta  
_{MCS}}},
\label{140}
\eeq
where ${\tilde \Delta _{MCS}} = \mid\mu\mid + 1$, 
and ${\tilde c _{MCS}} = \frac{\mid\mu\mid}{\pi}$.
Therefore $J _{i}^{MCS}$ has conformal dimension ${\tilde
\Delta _{MCS}}$. As it is well known the
Maxwell-Chern-Simons theory describes a particle with 
mass $\mu$ \cite{mcs} and
this fact is reflected in the conformal dimension ${\tilde \Delta _{MCS}}$.
Furthermore, our result is consistent with the holographic principle
since the mass $m_{-}(0,\mid\mu\mid) = 0$ is not physical in the bulk
\cite{mcs} and does not contribute to the border two-point function. 

For the self-dual model we start with the Euclidean signature action
\beq
I^0_{SD}=  \int d^3x \; \sqrt{g} \left( \frac{1} {\sqrt{g}}
  \frac{i\kappa}{8 } \epsilon^{\mu\nu\alpha} F_{\mu\nu}A_{\alpha} +
  \frac{1}{4} M^{2} A_{\mu} A^{\mu}\;+\;c.c. \right).
\label{145}
\eeq
In order to have
a stationary action we must supplement the action Eq.(\ref{145}) 
with a surface term which cancels its variation \cite{henneaux}. The
variational principle generates a boundary term 
\beq
-\frac{\kappa}{2}\int d^{2}x\;\epsilon^{0ij}
\;\left[ A_{i}^{R}(\vec{x})\delta
A_{j}^{I}(\vec{x})\;+\;A_{i}^{I}(\vec{x})\delta
A_{j}^{R}(\vec{x})\right], 
\label{ac2}
\eeq
which is written in terms of the real and imaginary parts of the vector
potential. Since the field equations derived from Eq.(\ref{145}) are
first order differential equations we can not choose boundary
conditions which fix simultaneously the real and imaginary parts of
the $A_i$'s. Then we choose boundary conditions on the $A_i^R$'s
leaving a non-vanishing term proportional to the $\delta A_i^I$'s in
the boundary term Eq.(\ref{ac2}). So we add to the action
Eq.(\ref{145}) a surface term of the form 
\beq
I^{surface}_{SD} = \frac{\kappa}{2}\int d^{2}x\;\epsilon^{0ij}
\;A_{i}^{R}(\vec{x})A_{j}^{I}(\vec{x}),
\label{ac5}
\eeq
and the action 
\beq
I_{SD} = I^{0}_{SD} + I^{surface}_{SD},
\label{ac4}
\eeq
is now stationary. 

The field equations which follow from the action Eq.(\ref{ac4}) are 
\beq
i \kappa \frac{1}{\sqrt{g}} \epsilon^{\nu\alpha\beta}
\partial_{\alpha}A_{\beta} + M^2 A^{\nu} = 0,
\label{1000}
\eeq
and it implies again 
\beq
\nabla_{\mu}A^{\mu}=0.
\label{1001}
\eeq
As in the case of the Proca-Chern-Simons theory we can eliminate the
Levi--Civita tensor density by increasing the order of the equations of
motion. We then get
\beq
\left( \nabla^2 - \frac{M^4}{\kappa^{2}} - \frac{R}{3} \right)
A^\mu = 0.
\label{1002}
\eeq
and the results of the Proca theory can again be used. 

The bulk to
bulk propagator can be calculated \cite{cs} and a regularization has
to be introduced
\beq
\lim_{\epsilon\to
0}\epsilon^{\frac{M^{2}}{\mid\kappa\mid} -
1}{\tilde A _{\epsilon,i}^{R}}(\vec{x}) = A _{0,i}(\vec{x}).
\label{identif3}
\eeq
Using the AdS/CFT correspondence Eq.(\ref{134}) we find the two-point
function of the 
conformal field $J_{i}^{SD}$ coupled to the field ${\tilde A_{i}}$ on the
boundary 
\beq
\left <J_{i}^{SD}(\vec{x})\; J_{j}^{SD}(\vec{y})\right> = 2
{\tilde c}_{SD}{\tilde
\Delta}_{SD}\left( \delta
_{ij} - 2\frac{(x-y) _{i}(x-y) _{j}}{\mid\vec{x} -
\vec{y}\mid^{2}}\right)\mid\vec{x} - \vec{y}\mid^{-2{\tilde \Delta}_{SD}},
\label{162}
\eeq
where $
{\tilde \Delta}_{SD} = \frac{M^{2}}{\mid\kappa\mid} + 1$, 
and $
{\tilde c}_{SD} = \frac{\mid\kappa\mid}{\pi}\;$. We then find that the
field $J_{i}^{SD}$ has conformal dimension ${\tilde 
\Delta}_{SD}$. Therefore 
the conformal dimensions of the conformal fields corresponding to the
Maxwell-Chern-Simons theory and the Self-Dual model are the same for
$\frac{M^{2}}{\mid\kappa\mid} = \mid\mu\mid$ in agreement with the
equivalence between those models \cite{equiv}. The fact that we
obtain the same conformal dimension for 
the corresponding CFT's in the border is in support of the
holographic principle. Not only the conformal dimensions are the
same but the coefficients $\tilde{c}$ of the two-point functions can
be made the same by an appropriate normalization of the Self-Dual
action. Since we started with two independent parameters in
Eq.(\ref{145}) we can now choose $M = | \kappa|$ so that the model
describes a particle with mass $M$. Now our results have an universal
form in which the conformal dimension and the two-point function
coefficient can be written as $\tilde{\Delta} = m +1$ and  $\tilde{c} =
m/\pi$ respectively, where $m$ is the mass of the bulk theory.  

\section{SCALAR FIELD THEORY}
\label{s3}

Scalar fields in AdS spaces have been intensively studied. If the
scalar field  has mass-squared in the 
range $-d^2/4<m^2<-d^2/4+1$ then there are two possible quantum field
theories in the bulk depending on the  choice of the energy-momentum
tensor \cite{freedman2}. The AdS/CFT correspondence with
Dirichlet boundary condition can easily
account for one of the theories. The other one appears in a very
subtle way by identifying a conjugate field through a Legendre
transform as the source of the boundary conformal operator \cite{witten2}.
The existence of two conjugated boundary operators has been first pointed
out in \cite{dobrev}.  

Since a field theory is determined not only by its
Lagrangian but also by its boundary terms we expect
that the AdS/CFT correspondence must be sensitive to these 
boundary terms. We will consider Dirichlet and Neumann boundary
conditions, and a 
combination of both of them which we will call mixed boundary
condition. 
Each type of boundary condition requires a different boundary term. We
will show that the mixed boundary conditions are parametrized by a real
number so that there is a one-parameter family of boundary terms
consistent with the variational principle. 
We will also show that different types of boundary condition give
rise to different conformal field theories at the border. 

The two solutions found in
\cite{freedman2} correspond to two different choices of 
energy-momentum tensor. For the Dirichlet boundary condition it is
well known that 
the scalar field behaves as $x_0^{d/2 - \sqrt{d^2/4+m^2}}$
near the border at $x_0=0$. There is no upper restriction on the mass in
this case. It corresponds to one of the solutions found in
\cite{freedman2} and gives rise to a boundary conformal operator
with conformal dimension  $d/2 + \sqrt{d^2/4+m^2}$. We will show that for
a particular
choice of mixed boundary condition and when the mass squared is in the
range $-d^2/4 < m^2 < -d^2/4+1 $ the scalar field behaves as
$x_0^{d/2 + \sqrt{d^2/4+m^2}}$ near the border. It corresponds
precisely to the second solution of 
\cite{freedman2} and gives rise to a boundary conformal operator
with conformal dimension  $d/2 - \sqrt{d^2/4+m^2}$. Note that the upper
limit for the mass squared $-d^2/4+1$ is consistent with the unitarity
bound $(d-2)/2$. Another important point that we will show is the
existence of boundary 
conditions which give rise to boundary conformal operators for which
the unitarity bound $(d-2)/2$ is 
reached. They correspond to a massless scalar field with Neumann
boundary condition or to a massive scalar field with $m^2 > -d^2/4 +1$
with a
particular choice of the mixed boundary condition (the same choice
which gives the boundary operator with conformal dimension
$d/2-\sqrt{d^2/4+m^2}$). In this way, using different boundary
conditions, we obtain all scalar conformal
field theories allowed by the unitarity bound. 

The action for the massive scalar field theory is given by 
\beq
I_{0} =  \frac{1}{2}\int d^{d+1}x \, \sqrt{g} \left(
g^{\mu\nu} \partial_{\mu}
\phi\partial_{\nu}\phi\;+\;m^{2}\phi^{2}\right),
\label{3}
\eeq 
and the corresponding equation of motion is
\beq
\left( \nabla^2 - m^2\right)\phi=0.
\label{4}
\eeq
In order to have a stationary action we must supplement the action 
$I_0$ with a boundary term $I_S$ which cancels its variation. The
appropriate action is then $I=I_{0} + I_{S}$. 

In order to capture the effect of the Minkowski boundary of the
$AdS_{d+1}$, situated at $x_0=0$, we first consider a boundary
value problem on the boundary surface $x_0=\epsilon > 0$ and
then take the limit $\epsilon \rightarrow 0$ at the very end. Then the
variational principle applied to the action $I$ gives 
\beq
\delta I = \int d^{d}x \;\epsilon^{-d+1} 
\;\partial_{0}\phi_{\epsilon}\;\delta\phi_{\epsilon} + \delta
I_{S} = 0, 
\label{8}
\eeq
where $\phi_\epsilon$ and $\partial_{0}\phi_{\epsilon}$ are the value
of the field and its derivate at $x_0=\epsilon$ respectively. 
This equation will be used below to find out the appropriate
boundary term $I_S$ for each type of boundary condition. 

For Dirichlet boundary condition the variation of the field at the
border vanishes so that the first term in Eq.(\ref{8}) also vanishes
and the usual action $I_{0}$ is already stationary.  
Making use of the field equation the action $I$ takes the form 
\beq
I_{D}= \frac{1}{2}\int
d^{d+1}x \; \partial_{\mu}\left(
\sqrt{g}\;\phi\;\partial^{\mu}\phi\right)=-\frac{1}{2}\int d^{d}x
\;\epsilon^{-d+1}
\;\phi_{\epsilon}\;\partial_{0}\phi_{\epsilon}\;.
\label{11a}
\eeq
It is to be understood that
$\partial_{0}\phi_{\epsilon}$ in Eq.(\ref{11a}) is evaluated in terms
of the Dirichlet data $\phi_{\epsilon}$. 

To consider Neumann boundary conditions we first take a unitary
vector which is inward normal to the boundary $
n^{\mu}(x_{0})=(x_{0},{\bf 0})$. The Neumann boundary condition then
fixes the value of $n^{\mu}(\epsilon) \partial_{\mu}\phi_\epsilon
\equiv \partial_n \phi_\epsilon$. The boundary term to be added to the
action reads
\beq
I_S= - \int
d^{d+1}x \; \partial_{\mu}\left(
\sqrt{g}\; g^{\mu\nu} \; \phi\;\partial_{\nu}\phi\right)=\int d^{d}x
\;\epsilon^{-d+1} 
\;\phi_{\epsilon}\;\partial_{0}\phi_{\epsilon}\;,
\label{9'}
\eeq
so that we find the following
expression for the action at the boundary
\beq
I_{N}=
\frac{1}{2}\int d^{d}x
\;\epsilon^{-d}
\;\phi_{\epsilon}\;\partial_{n}\phi_{\epsilon}\;.
\label{12}
\eeq
Here $\phi_{\epsilon}$ is to be expressed in terms of the Neumann
value $\partial_{n}\phi_{\epsilon}$. 
Notice that the 
on-shell value of the action with Neumann boundary condition
Eq.(\ref{12}) differs by a sign from the corresponding action with
Dirichlet boundary condition Eq.(\ref{11a}). 

We now consider a boundary condition which fixes the value of a
linear combination of the field and its normal derivative at the border
\beq
\phi(x) + \alpha n^{\mu}\partial_{\mu}\phi(x) \equiv \psi^{\alpha}(x)
\; . 
\label{100}
\eeq
We will call it mixed boundary condition. Here $\alpha$ is an arbitrary 
real but non-zero
coefficient. In this case the surface term to be added to the action
is 
\beq
I_{S}^{\alpha} = \frac{\alpha}{2}\int
d^{d+1}x \; \partial_{\mu}\left(
\sqrt{g}\; g^{\mu\nu} \partial_{\nu}\phi\; n^\rho \partial_{\rho}\phi
\right)=
-\frac{\alpha}{2}\int d^{d}x\;\epsilon^{-d+2}
\;\partial_{0}\phi_{\epsilon}\;\partial_{0}\phi_{\epsilon}\;,
\label{102}
\eeq
and we find the following expression for the action at the boundary
\beq
I_{M}^{\alpha} = -\frac{1}{2}\int d^{d}x
\;\epsilon^{-d+1}
\;\psi^{\alpha}_{\epsilon}\;\partial_{0}\phi_{\epsilon}\;.
\label{104}
\eeq
Clearly $\partial_{0}\phi_{\epsilon}$ in the above expression must be
written in terms of the boundary data $\psi^{\alpha}_{\epsilon}$. 
We then have a one-parameter family of surface terms since the
variational 
principle does not impose any condition on $\alpha$. In this way the
value of the on-shell action Eq.(\ref{104}) also depends on $\alpha$. 

We can now calculate the bulk to bulk propagator for each boundary
condition. The details are given in \cite{revisited}. We then find the
following two-point functions. For
Dirichlet boundary conditions:
\ba
&&\left<{\cal O}_{D}^{\nu\not=0}(\vec{x}){\cal
O}_{D}^{\nu\not=0}(\vec{y})\right>=\frac{2\nu}{\pi^{\frac{d}{2}}}\;
\frac{\Gamma(\frac{d}{2}+\nu)}{\Gamma(\nu)}\;\frac{1}{|\;
\vec{x}-\vec{y}\;|^{2(\frac{d}{2}+\nu)}}
\label{260}\\
&&\left<{\cal O}_{D}^{\nu=\frac{d}{2}}(\vec{x}){\cal
O}_{D}^{\nu=\frac{d}{2}}(\vec{y})\right>=\frac{d}{\pi^{\frac{d}{2}}}\;
\frac{\Gamma(d)}{\Gamma(\frac{d}{2})}\;\frac{1}{|\;
\vec{x}-\vec{y}\;|^{2d}}
\label{261}\\
&&\left<{\cal O}_{D}^{\nu=0}(\vec{x}){\cal
O}_{D}^{\nu=0}(\vec{y})\right>=
\frac{\Gamma\left(\frac{d}{2}\right)}{2\pi^{\frac{d}{2}}}\;
\frac{1}{|\;
\vec{x}-\vec{y}\;|^{d}}
\ea
\label{262}

For Neumann boundary conditions:
\ba
&&\left<{\cal O}_{N}^{\nu\not=0,\frac{d}{2}}(\vec{x}){\cal
O}_{N}^{\nu\not=0,\frac{d}{2}}(\vec{y})\right>=
\frac{1}{\sigma^{2}(\nu)}\;
\left<{\cal O}_{D}^{\nu\not=0}(\vec{x}){\cal
O}_{D}^{\nu\not=0}(\vec{y})\right>
\label{263}\\
&&\left<{\cal O}_{N}^{\nu=\frac{d}{2}}(\vec{x}){\cal
O}_{N}^{\nu=\frac{d}{2}}(\vec{y})\right>=
\frac{\Gamma(\frac{d}{2})}{2\pi^{\frac{d}{2}}}\;\frac{1}{|\;
\vec{x}-\vec{y}\;|^{2\frac{d-2}{2}}}
\label{264}\\
&&\left<{\cal O}_{N}^{\nu=0}(\vec{x}){\cal
O}_{N}^{\nu=0}(\vec{y})\right>=
\frac{1}{\sigma^{2}(0)}\;\left<{\cal O}_{D}^{\nu=0}(\vec{x}){\cal
O}_{D}^{\nu=0}(\vec{y})\right>
\label{265}
\ea

And for mixed boundary conditions:
\ba
&&\left<{\cal O}_{M}^{\beta=0,0<\nu<1}(\vec{x}){\cal
O}_{M}^{\beta=0,0<\nu<1}(\vec{y})\right>=\sigma^{2}(\nu)
\frac{1}{2\pi^{\frac{d}{2}}}\;  
\;\frac{\Gamma(\frac{d}{2}-\nu)}{\Gamma(1-\nu)}\;\frac{1}{|\;
\vec{x}-\vec{y}\;|^{2(\frac{d}{2}-\nu)}}
\label{360}\\
&&\left<{\cal O}_{M}^{\beta=0,\nu >1}(\vec{x}){\cal
O}_{M}^{\beta=0,\nu >1}(\vec{y})\right>=\sigma^{2}(\nu)\;(\nu-1)\;
\frac{\Gamma\left( \frac{d-2}{2}\right)}{2\pi^{\frac{d}{2}}}\;
\frac{1}{|\;  
\vec{x}-\vec{y}\;|^{2\frac{d-2}{2}}}
\label{361}\\
&&\left<{\cal
O}_{M}^{\beta\not=0,\nu\not=0,\frac{d}{2}}(\vec{x}){\cal
O}_{M}^{\beta\not=0,\nu\not=0,\frac{d}{2}}(\vec{y})\right>=
\frac{1}{\beta^{2}(\alpha,\nu)}\;\left<{\cal O}_{D}^{\nu\not=0}(\vec{x}){\cal
O}_{D}^{\nu\not=0}(\vec{y})\right>
\label{719}\\
&&\left<{\cal O}_{M}^{\nu=\frac{d}{2}}(\vec{x}){\cal
O}_{M}^{\nu=\frac{d}{2}}(\vec{y})\right>=\left<{\cal
O}_{D}^{\nu=\frac{d}{2}}(\vec{x}){\cal
O}_{D}^{\nu=\frac{d}{2}}(\vec{y})\right>
\label{820}\\
&&\left<{\cal O}_{M}^{\beta\not=0,\nu=0}(\vec{x}){\cal
O}_{M}^{\beta\not=0,\nu=0}(\vec{y})\right>=\frac{1}{\beta^{2}(\alpha,0)}
\;\left<{\cal
O}_{D}^{\nu=0}(\vec{x}){\cal
O}_{D}^{\nu=0}(\vec{y})\right>
\label{721}\\
&&\left<{\cal O}_{M}^{\beta=0,\nu=0}(\vec{x}){\cal
O}_{M}^{\beta=0,\nu=0}(\vec{y})\right>=\sigma^{2}(0)
\;\left<{\cal
O}_{D}^{\nu=0}(\vec{x}){\cal
O}_{D}^{\nu=0}(\vec{y})\right>
\label{722}
\ea

Here the coefficients $\nu$, $\beta$ and $\sigma$ are defined
as $\nu = \sqrt{\frac{d^{2}}{4}+m^{2}}\;$,
$\beta(\alpha,\nu) = 1+\alpha\left( \frac{d}{2}-\nu\right)$ and
$\sigma (\nu) = d/2 - \nu$ respectively. 

We then find that different boundary conditions in the AdS/CFT
correspondence allow us to derive boundary two-point functions which are
consistent with the unitarity bound. In the Neumann case the unitarity
bound is 
obtained for $m=0$ while with mixed boundary conditions it is reached
when $\beta=0$ and $m^{2}>-d^{2}/4 + 1$. The corresponding two-point
functions have different normalizations. The conformal dimension $d/2 -
\nu$ is obtained in the case of mixed boundary condition with $\beta=0$
and $-d^{2}/4<m^{2}<-d^{2}/4 + 1$, and the normalization of
the corresponding boundary two-point function differs from the one found
in \cite{witten2}. 

Another important point is the interpretation of the new boundary
conditions in the string theory context. Dirichlet boundary conditions are
natural when thinking of the asymptotic behavior of the supergravity
fields reaching the border of the AdS space. Possibly Neumann and mixed
boundary conditions are related to more complex solutions involving
strings and membranes reaching the border in more subtle ways.

\section{ABELIAN GAUGE THEORY}
\label{s4}

For the Abelian gauge field in a fixed gauge the corresponding
conformal correlators were found in \cite{ads/cft}. Now we will
consider the gauge dependence of this result. To do that we take the 
formulation of electrodynamics in an arbitrary
gauge when the space-time background is $AdS_{d+1}$. As in the case of
flat Minkowski space it will prove convenient to start from the St\"uckelberg
action 
\be
\label{102a}
I_S\,=\,\,\int\,d^{d+1}x\,{\sqrt g}\,\left[\,\frac{1}{4}\,F_{\alpha \beta}\,
g^{\alpha \rho}\,g^{\beta \sigma}\,F_{\rho \sigma}\,+\,\frac{m^2}{2}\,
A_{\alpha}\,g^{\alpha \beta}\,A_{\beta}\,+\,\frac{1}{2a}\,\left(g^{\alpha
\beta}\, \nabla_{\alpha} A_{\beta}\right)^2 \right]\,\,\,,
\ee
where $F_{\alpha \beta} = \pa_{\alpha}A_{\beta} - \pa_{\beta}A_{\alpha}$,
$\nabla_{\alpha}$ is the covariant derivative and $a$ is a real 
positive constant. Electrodynamics in an arbitrary
covariant gauge, specified by the constant $a$, is defined as the limit 
$m^2 \rightarrow 0$ of  St\"uckelberg theory. On the other hand the limit 
$a \rightarrow \infty$, while keeping $m^2 > 0$, results in the Proca theory.
The mass term in (\ref{102a}) will help us to control the infrared divergent 
terms which will arise along the calculation. 

The Lagrange equations of motion arising from (\ref{102a}) are 
\be
\label{1031}
\nabla_{\mu} F^{\mu \nu}\,+\,\frac{1}{a}\,\nabla_{\mu} L^{\mu \nu}\,-\,m^2
A^{\nu} \,=\,0\,\,\,,
\ee
where $L^{\mu \nu} \equiv g^{\mu \nu} g^{\alpha \beta} \nabla_{\alpha}
A_{\beta}$. We now use a decomposition of  $A^{\nu}$ into a
scalar field $\Phi \equiv\,\nabla_{\nu} A^{\nu}$, and a vector field
$U^{\nu}\,\equiv\,A^{\nu}\,-\,\frac{1}{a m^2}\,\nabla^{\nu}\Phi$. 
These new fields satisfy, respectively, the equations of motion
\be
\label{106}
\left(\,g_{\mu \nu} \nabla^{\mu} \nabla^{\nu}\,-\,a m^2\,\right)\Phi\,=\,0,
\ee
and
\be
\label{107}
\nabla_{\mu} U^{\mu \nu}\,-\,m^2 U^{\nu}\,=\,0\,\,\,.
\ee
Clearly $U^{\mu}$ is a 
Proca field with mass $m$ since $\nabla_{\mu} U^{\mu}\,=\,0$.

As usual we shall look for a solution written in terms of boundary field
values specified at the near-boundary surface $x^0=\epsilon$ the limit 
$\epsilon \rightarrow 0$ being performed at the very end of the
calculations. The details are given in \cite{gauge}. First of all we
must make sure that the massless limit can be taken. There are
infrared divergences in $\Phi/am^2$ and $U_{\nu}$ but a careful
analysis shows that they cancel out when the field $A_\mu$ is
recomposed. This shows that $A_\mu$ is indeed an analytic function of
$m^2$ in the vicinity of $m^2 = 0$. 

In the computation of the two-point function we find that all
potentially dangerous powers of $\epsilon$ cancel out among
themselves. We then find that the gauge dependence concentrates on the
contact terms while the non-trivial part of the boundary conformal theory 
correlators turns out to be that already found by working in a
completely fixed gauge\cite{ads/cft}. 
Another important feature is that although we have
fixed all components of the potential at the border the pieces
containing $\tilde{A}_{\epsilon,0}$ give only contact terms and the only
non-trivial pieces are those containing $\tilde{A}_{\epsilon,i}$. Therefore
the boundary theory still retains information on the gauge degrees of
freedom of the bulk theory. This then lends further support to the
holographic principle.

\section{ACKNOWLEDGEMENTS}

This work is partially supported by CNPq and FAPESP.

\end{document}